\documentclass[3p,times,procedia]{elsarticle}
\usepackage{nupha_ecrc}

\volume{00}

\firstpage{1}

\journalname{Nuclear Physics A}

\runauth{E. Scomparin}

\jid{nupha}

\jnltitlelogo{Nuclear Physics A}

\usepackage{amssymb}

\usepackage[figuresright]{rotating}

\begin{document}

\begin{frontmatter}



\dochead{XXVIth International Conference on Ultrarelativistic Nucleus-Nucleus Collisions\\ (Quark Matter 2017)}

\title{Quarkonium production in \mbox{A-A} and \mbox{p-A} collisions}


\author{E. Scomparin}

\address{Istituto Nazionale di Fisica Nucleare -- Sezione di Torino, Via Giuria 1, 10125 Torino, Italy}

\begin{abstract}
Thirty years ago, the suppression of quarkonium production in heavy-ion collisions was first proposed as an unambiguous signature for the formation of a Quark-Gluon Plasma. Recent results from the LHC run 2 have led to an unprecedented level of precision on this observable and, together with new data  from RHIC, are providing an accurate picture of the influence of the medium created in nuclear collisions on the various charmonium (J/$\psi$, $\psi$(2S)) and bottomonium ($\Upsilon(1S)$, $\Upsilon(2S)$, $\Upsilon(3S)$) states, studied via  their decay into lepton pairs.
In this contribution, I will review the new results presented at Quark Matter 2017, emphasizing their relation with previous experimental observations and comparing them, where possible, with theoretical calculations.
\end{abstract}


\end{frontmatter}


\section{Introduction}
\label{sec:introduction}

In August 1987, at the 6$^{\rm th}$ Quark Matter Conference in Nordkirchen (Germany), the NA38 Collaboration presented the first evidence for a suppression of the J/$\psi$ in \mbox{O-U} collisions at 200 GeV/nucleon~\cite{Abreu:1988tp}. By showing that the ratio between the number of J/$\psi$ decaying to muon pairs and the corresponding number of continuum events in  the same mass region was decreasing by $\sim$35\% between peripheral and central events, charmonium suppression became the ``smoking gun'' for the observation of deconfinement, a status retained to a considerable extent also in more recent times. However, it became clear rather soon that the interpretation of the experimental observations was far from being trivial, due to competing suppression sources, which included cold nuclear matter effects and dissociation processes in a hot hadronic gas.

After thirty years, and several rounds of experiments, starting with the fixed target program at the CERN SPS and then moving to colliders (RHIC and LHC), a consensus has been reached on the qualitative interpretation of the charmonium suppression results in terms of a superposition of color screening and recombination mechanisms. The first conceptually derives from the Matsui and Satz original prediction~\cite{Matsui:1986dk}, while the recombination effect~\cite{BraunMunzinger:2000px,Thews:2000rj}, which tends to counterbalance color screening, becomes evident at the LHC energy where the multiplicity of $c\overline c$ pairs is very large ($>100$ for central \mbox{Pb-Pb} collisions). The most evident manifestation of the strong contribution of recombination processes is visible in the comparison of the J/$\psi$ nuclear modification factors ($R_{\rm AA}$) between RHIC and LHC energy, with systematically larger $R_{\rm AA}$ values at the LHC, and in particular for low transverse momentum ($p_{\rm T}$) values~\cite{Adam:2015isa}.
For bottomonium, even at the LHC the recombination effects are expected to be marginal, and a hyerarchy of suppression was observed in run 1 LHC results, with smaller $R_{\rm AA}$ values for the more weakly bound $\Upsilon$ states~\cite{Khachatryan:2016xxp}.

Although the aforementioned mechanisms can qualitatively explain the observations, a satisfactory quantitative description is far from being trivial.
In particular, aspects such as the spectral modifications of quarkonium states in hot matter and their formation time compared to that of QGP play an important role and are presently much debated in the theory community. On the experimental side, it is clear that the availability of more accurate data, in particular for what concerns excited quarkonium states and/or differential studies of $R_{\rm AA}$ can help to constrain the model calculations. Connected to that, a precise determination of the open charm cross section, to calibrate the size of the recombination process, is also needed.

In the next Sections I will review the recent experimental achievements on quarkonium production in nuclear collisions, with an emphasis on results presented at the Quark Matter 2017 Conference and discussing a personal choice of the main highlights.

\section{Charmonium production: A-A collisions}
\label{sec:charmAA}

\begin{figure}[hbtp]
\begin{center}
\includegraphics[width=0.48\linewidth]{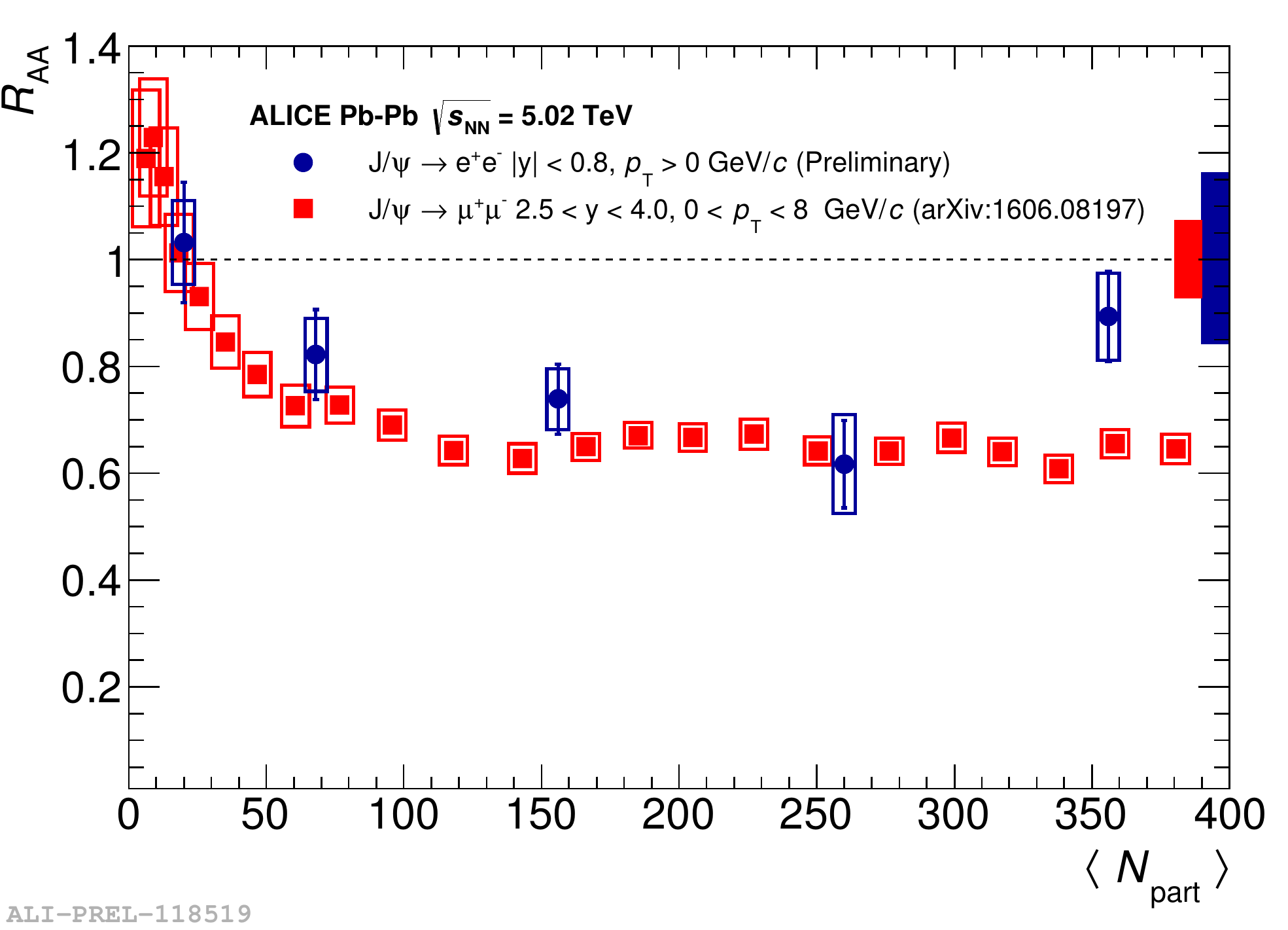} 
\includegraphics[width=0.48\linewidth]{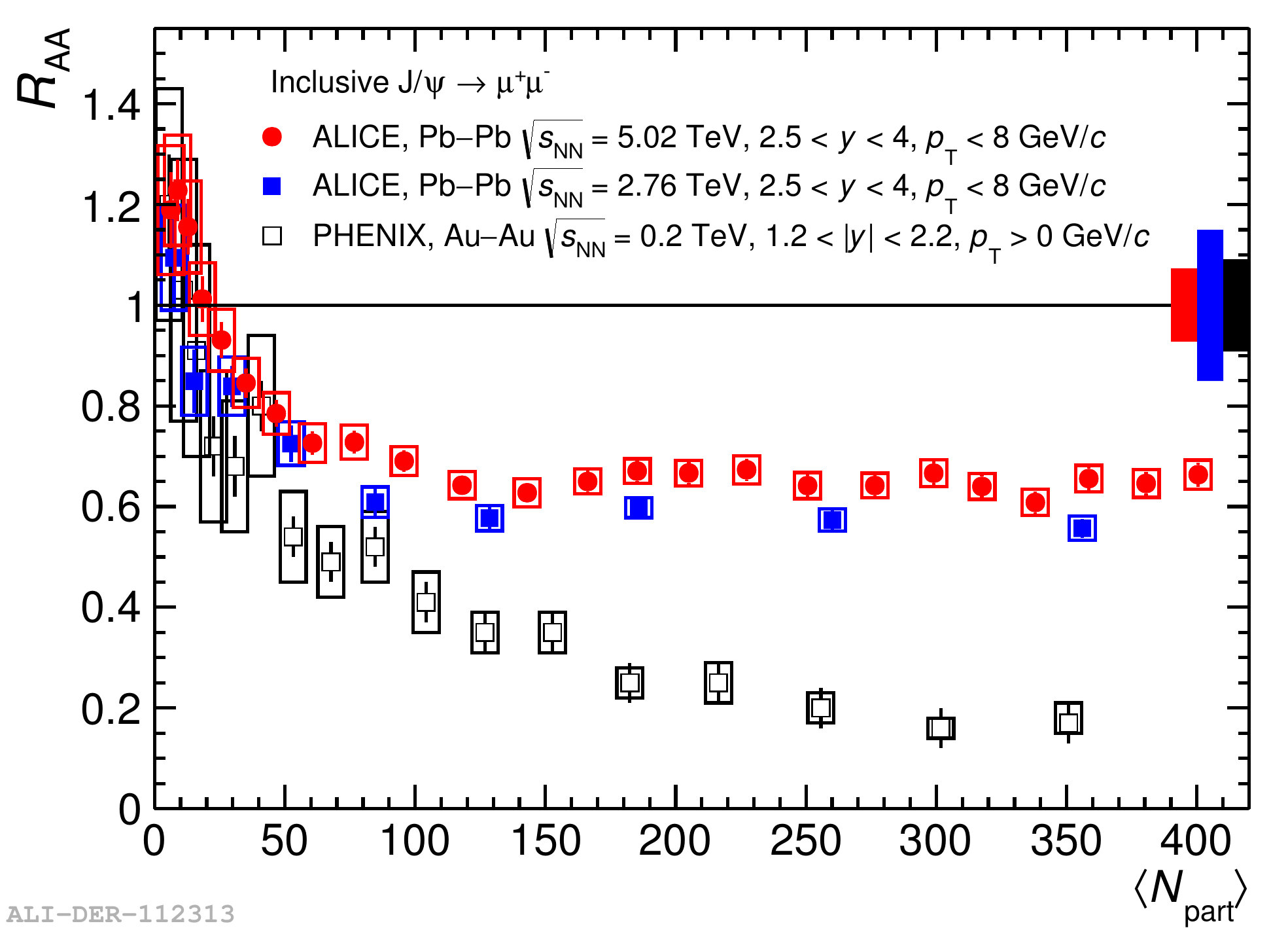} 
\caption{Left: centrality dependence of the inclusive J/$\psi$ $R_{\rm AA}$ for \mbox{Pb-Pb} collisions at $\sqrt{s_{\rm NN}}=5.02$ TeV, measured by ALICE, in two $y$ regions~\cite{QM17:Bustamante,Adam:2016rdg}. Right: inclusive J/$\psi$ $R_{\rm AA}$ as a function of centrality for various $\sqrt{s_{\rm NN}}$ values, from ALICE and PHENIX~\cite{Adam:2015isa,Adam:2016rdg,Adare:2011yf}.}
\label{fig:1}
\end{center}
\end{figure}

In \mbox{Pb-Pb} collisions, inclusive J/$\psi$ production results were presented by the ALICE Collaboration, covering the forward ($2.5<y<4$) and central ($|y|<0.9$) rapidity regions~\cite{QM17:Tarhini,QM17:Bustamante}. Since the kinematic coverage extends to $p_{\rm T}=0$, the centrality dependence of $R_{\rm AA}$ is dominated by low-$p_{\rm T}$ J/$\psi$, where recombination effects are expected to be stronger. In Fig.~\ref{fig:1}~(left) the preliminary results for the inclusive J/$\psi$ $R_{\rm AA}$ at central rapidity in \mbox{Pb-Pb} collisions at $\sqrt{s_{\rm NN}}=5.02$ TeV~\cite{QM17:Bustamante}  are compared with the recently published forward-$y$ results~\cite{Adam:2016rdg}, as a function of centrality. The latter, thanks to the very fine centrality binning, show a clear saturation of the J/$\psi$ suppression for values of the number of participant nucleons $N_{\rm part}>100$. The midrapidity result exhibits the same feature with a hint for a weaker suppression for very central events. When comparing, as in Fig.~\ref{fig:1}~(right), the forward-$y$ results with the corresponding ones from the LHC run 1 ($\sqrt{s_{\rm NN}}=2.76$ TeV)~\cite{Adam:2015isa}, agreement is found within uncertainties, an observation likely indicating that any additional suppression due to the larger energy density at higher collision energy is counterbalanced by a stronger recombination effect. The stronger suppression visible in PHENIX data~\cite{Adare:2011yf} clearly implies that recombination plays a less important role at the lower collision energy.

\begin{figure}[hbtp]
\begin{center}
\includegraphics[width=0.5\linewidth]{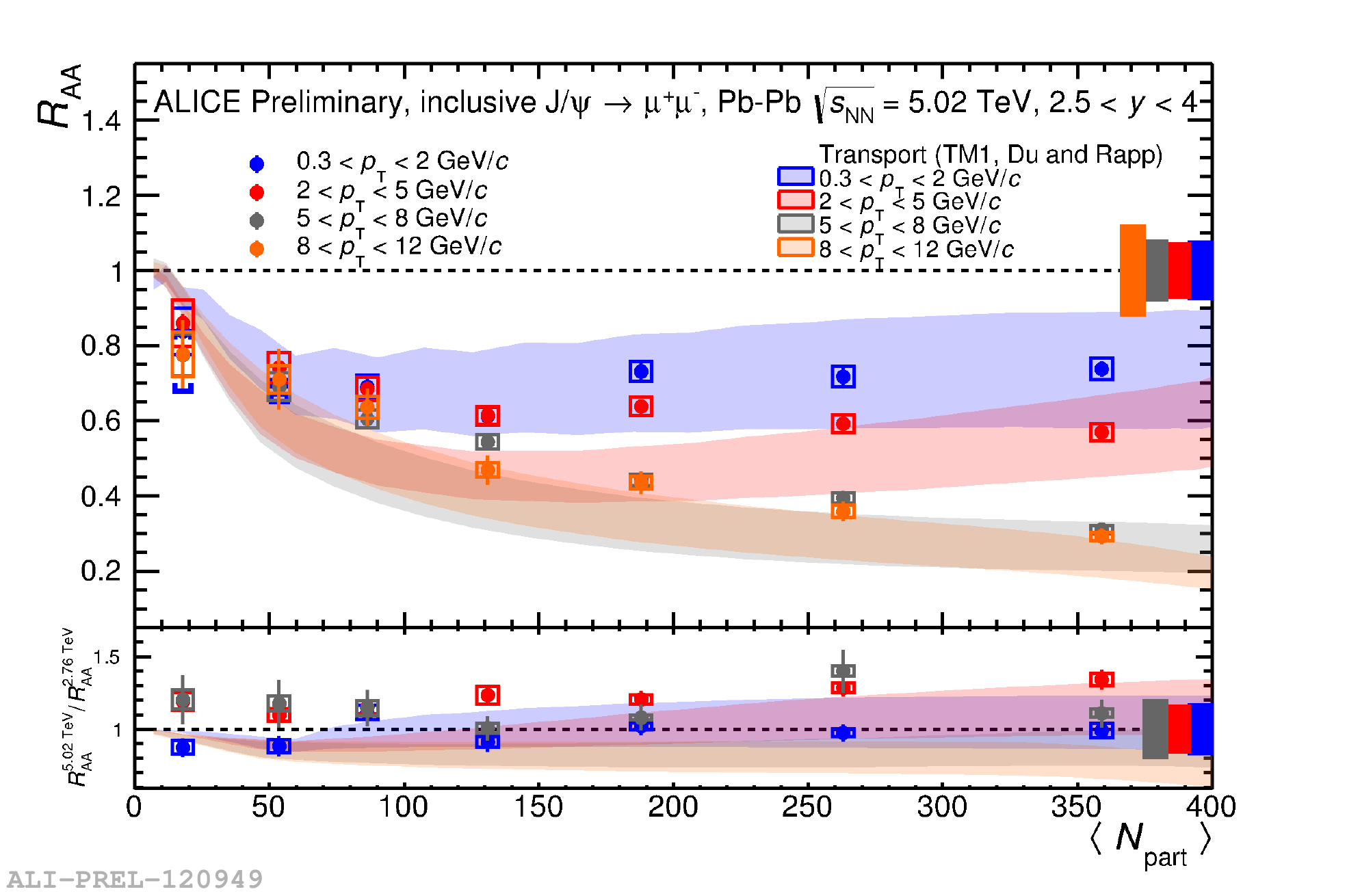} 
\includegraphics[width=0.47\linewidth]{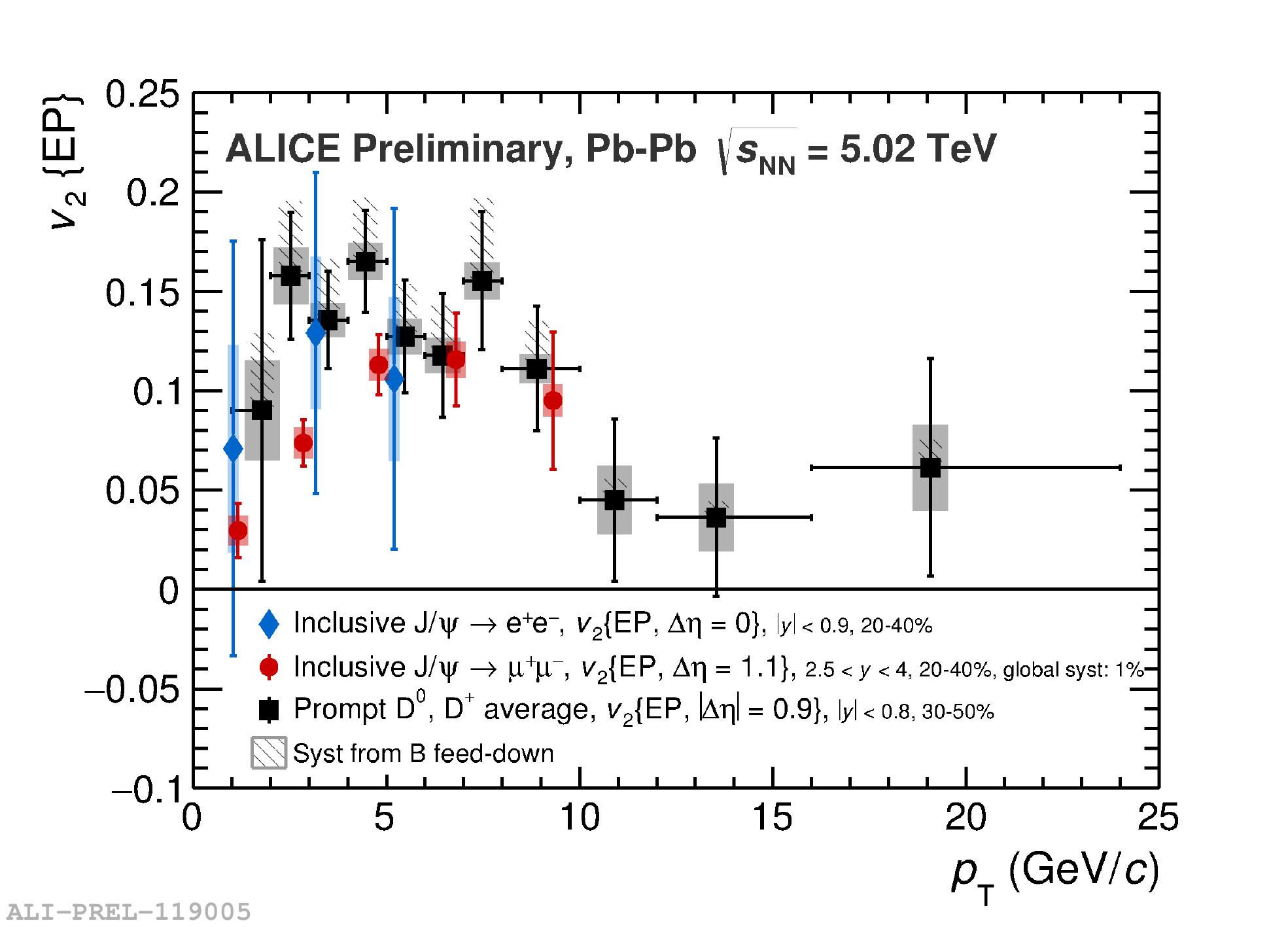} 
\caption{Left: centrality dependence of the inclusive J/$\psi$ $R_{\rm AA}$ for \mbox{Pb-Pb} collisions at $\sqrt{s_{\rm NN}}=5.02$ TeV, measured by ALICE for $2.5<y<4$ and in four $p_{\rm T}$ bins~\cite{QM17:Tarhini}, compared with the result of a theory calculation~\cite{Zhao:2011cv}. Right: the $p_{\rm T}$-dependence of inclusive J/$\psi$ $v_2$ for 20-40\% \mbox{Pb-Pb} collisions~\cite{QM17:Tarhini}, measured by ALICE in two rapidity ranges, and compared with the D-meson $v_2$ for 30-50\% centrality~\cite{QM17:Barbano}.}
\label{fig:2}
\end{center}
\end{figure}

Thanks to the large integrated luminosity for the $\sqrt{s_{\rm NN}}=5.02$ TeV data sample, it is now possible to study the centrality dependence of the $R_{\rm AA}$ differentially as a function of the J/$\psi$ kinematic variables~\cite{QM17:Tarhini}. In Fig.~\ref{fig:2}~(left), a new result from ALICE on the nuclear modification factor is presented, for four $p_{\rm T}$ bins, as a function of $\langle N_{\rm part}\rangle$. While for peripheral collisions no significant dependence on $p_{\rm T}$ is visible, for central collisions a strong reduction of the suppression can be seen at low $p_{\rm T}$, making the centrality dependence of $R_{\rm AA}$ in $0.3<p_{\rm T}<2$ GeV/$c$ almost flat. The comparisons with a theoretical model~\cite{Zhao:2011cv} shows that the main features of the data can be reproduced, but some tension persists, e.g., at intermediate $p_{\rm T}$ and centrality. A systematic comparison of such multi-differential $R_{\rm AA}$ measurements with theoretical calculations represents a promising way to pin down quantitatively the features of the various mechanisms at play.

As a further step forward, the first observation of a non-zero elliptic flow  ($v_2$) for the J/$\psi$ in \mbox{Pb-Pb} collisions at $\sqrt{s_{\rm NN}}=5.02$ TeV was presented by ALICE at the Conference~\cite{QM17:Tarhini}. In Fig.~\ref{fig:2}~(right), the $p_{\rm T}$ dependence of $v_2$ is shown for the 20-40\% centrality interval, in the ranges $|y|<0.9$ and $2.5<y<4$. A significance up to 7.4$\sigma$ for a non-zero $v_2$ was estimated for the forward-$y$ data. This result may indicate that a significant fraction of the observed J/$\psi$ comes from charm quarks which thermalized in the QGP. The comparison, also shown in Fig.~\ref{fig:2}~(right), with new results on the D-meson elliptic flow in the 30-50\% centrality range~\cite{QM17:Barbano}, opens up the possibility of investigating the relation between light and heavy quark flow in the medium.

\begin{figure}[hbtp]
\begin{center}
\includegraphics[width=0.42\linewidth]{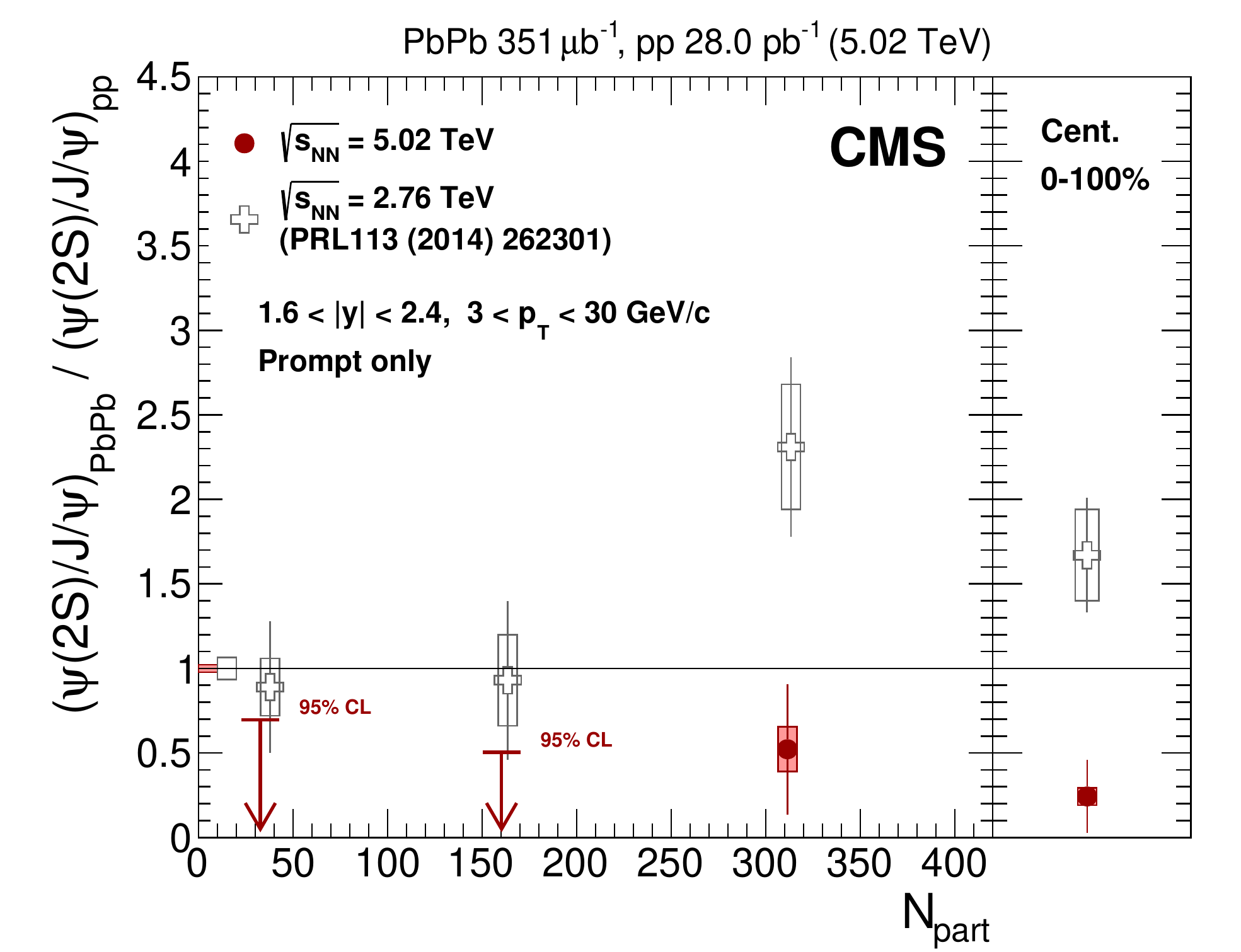} 
\includegraphics[width=0.51\linewidth]{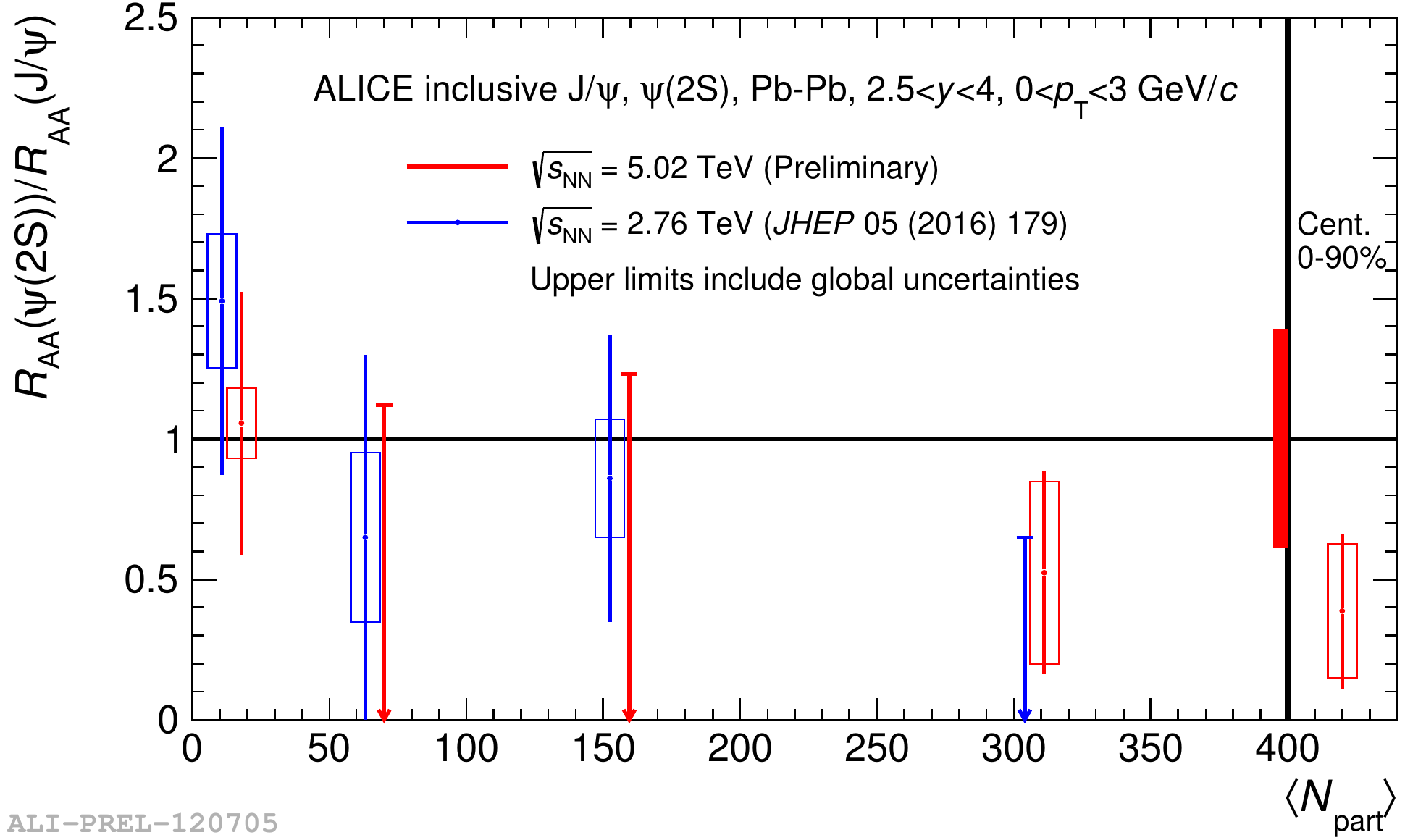} 
\caption{Left: centrality dependence of the double ratio of $\psi(2S)$ and J/$\psi$ cross sections between \mbox{Pb-Pb} and pp, measured by CMS at $\sqrt{s_{\rm NN}}=2.76$ and 5.02 TeV for $3<p_{\rm T}<30$ GeV/$c$~\cite{Sirunyan:2016znt}. 
Right: same quantity, measured by ALICE for $p_{\rm T}<3$ GeV/$c$~\cite{QM17:Tarhini}.}
\label{fig:3}
\end{center}
\end{figure}

The study of the weakly bound $\psi(2S)$ represents a potentially important tool, as one would expect much stronger dissociation effects in the QGP, while the evaluation of recombination is less straightforward, as it might become  important at a later stage in the collision history. Recent results from CMS~\cite{QM17:MartinBlanco,Sirunyan:2016znt}, shown in Fig.~\ref{fig:3}~(left), indicate for \mbox{Pb-Pb} collisions at $\sqrt{s_{\rm NN}}=5.02$ TeV, a stronger suppression for the $\psi(2S)$ with respect to the J/$\psi$, quantified through the measurement of the double ratio
of $\psi(2S)$ to J/$\psi$ cross sections between \mbox{Pb-Pb} and pp. This result, which refers to the region $1.6<|y|<2.4$, $3<p_{\rm T}<30$ GeV/$c$, shows a different trend with respect to the corresponding one obtained at $\sqrt{s_{\rm NN}}=2.76$ TeV~\cite{Khachatryan:2014bva}, where a strong enhancement of the double ratio was observed, in particular for central events. Preliminary ALICE results~\cite{QM17:Tarhini}, covering the region down to zero $p_{\rm T}$, were shown at the Conference (Fig.~\ref{fig:3}~(right)). In particular, for $p_{\rm T}<3$ GeV/$c$ and $2.5<y<4$, values of the double ratio smaller than 1 were measured. The only proposed mechanism for the enhancement shown by CMS at $\sqrt{s_{\rm NN}}=2.76$ TeV involves a $\psi(2S)$ regeneration mechanism occurring late in the collision history, when radial flow is already built-up and therefore the charmonia have acquired a significant transverse momentum~\cite{Du:2016fam}. Also in this case, it will be important to provide systematic comparisons between data and calculations to validate this picture.

\section{Charmonium production: p-A collisions}
\label{sec:charmpA}

In the context of charmonium production, the study of \mbox{p-A} collisions is relevant for the interpretation of \mbox{A-A} results, since cold nuclear matter effects, which can be studied via such collisions and are not related to the production of a QGP, are known to be sizeable. In particular, parton shadowing in the nucleus, coherent parton energy loss and saturation phenomena related to the formation of a Color Glass Condensate (CGC) are also interesting from a more general point of view, to investigate aspects of QCD that are up to now not completely settled. Studies of \mbox{p-Pb} collisions during the LHC run 1 ($\sqrt{s_{\rm NN}}=5.02$ TeV) have shown that J/$\psi$ nuclear modification factors as low as $R_{\rm pPb}\sim 0.6$ can be reached at forward rapidity ($y_{\rm cms}\sim 3.5$) and low $p_{\rm T}$, a value similar to the one observed in \mbox{Pb-Pb}~\cite{Adam:2015iga}. In other words, this implies that by  correcting the measured $R_{\rm AA}$ for cold nuclear matter effects,
one would end up with $R_{\rm AA}\sim 1$ at low $p_{\rm T}$, indicating that suppression and recombination effects for low-$p_{\rm T}$ J/$\psi$ in \mbox{Pb-Pb} are of a similar size.

\begin{figure}[hbtp]
\begin{center}
\includegraphics[width=0.48\linewidth]{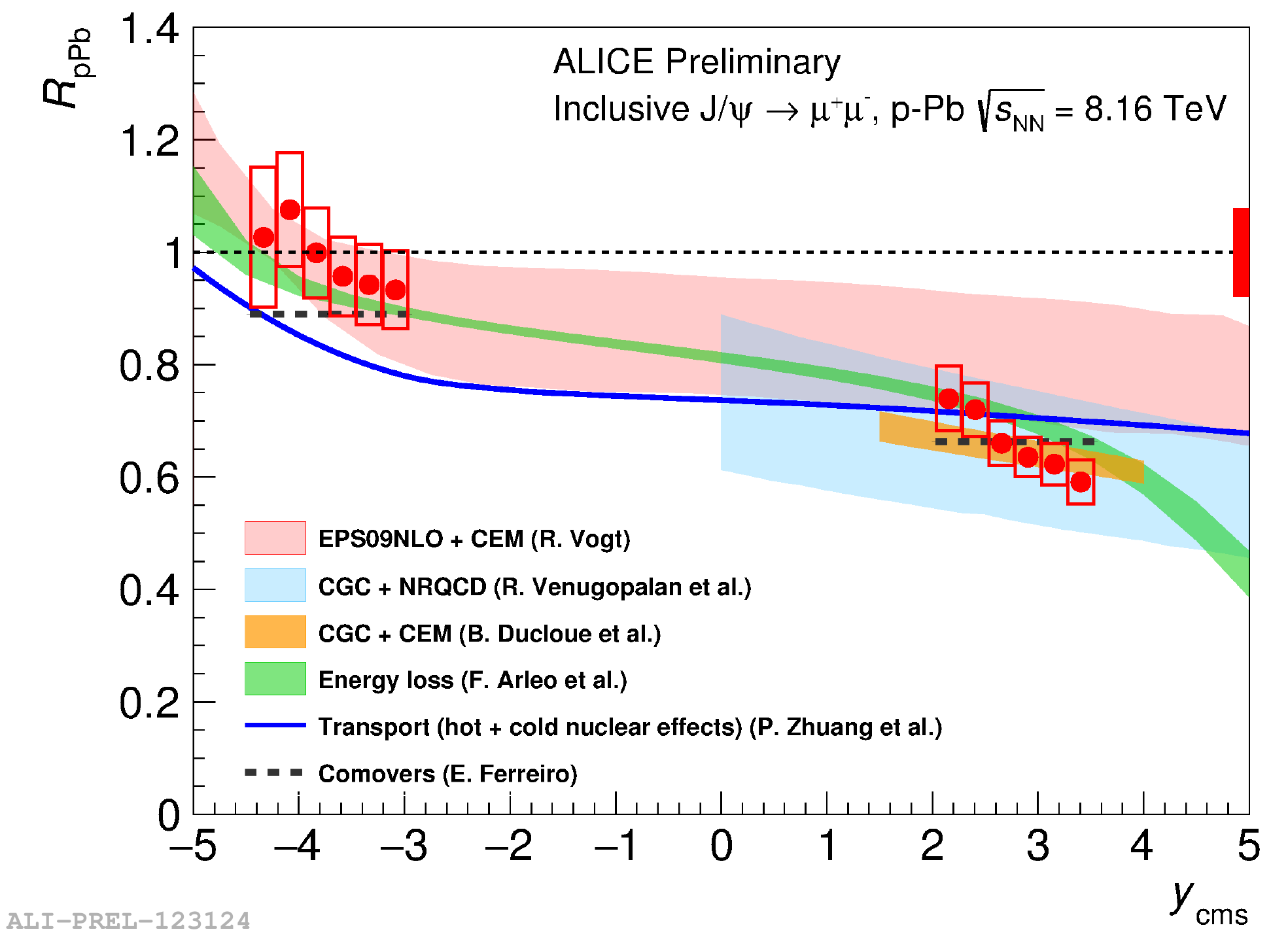} 
\includegraphics[width=0.48\linewidth]{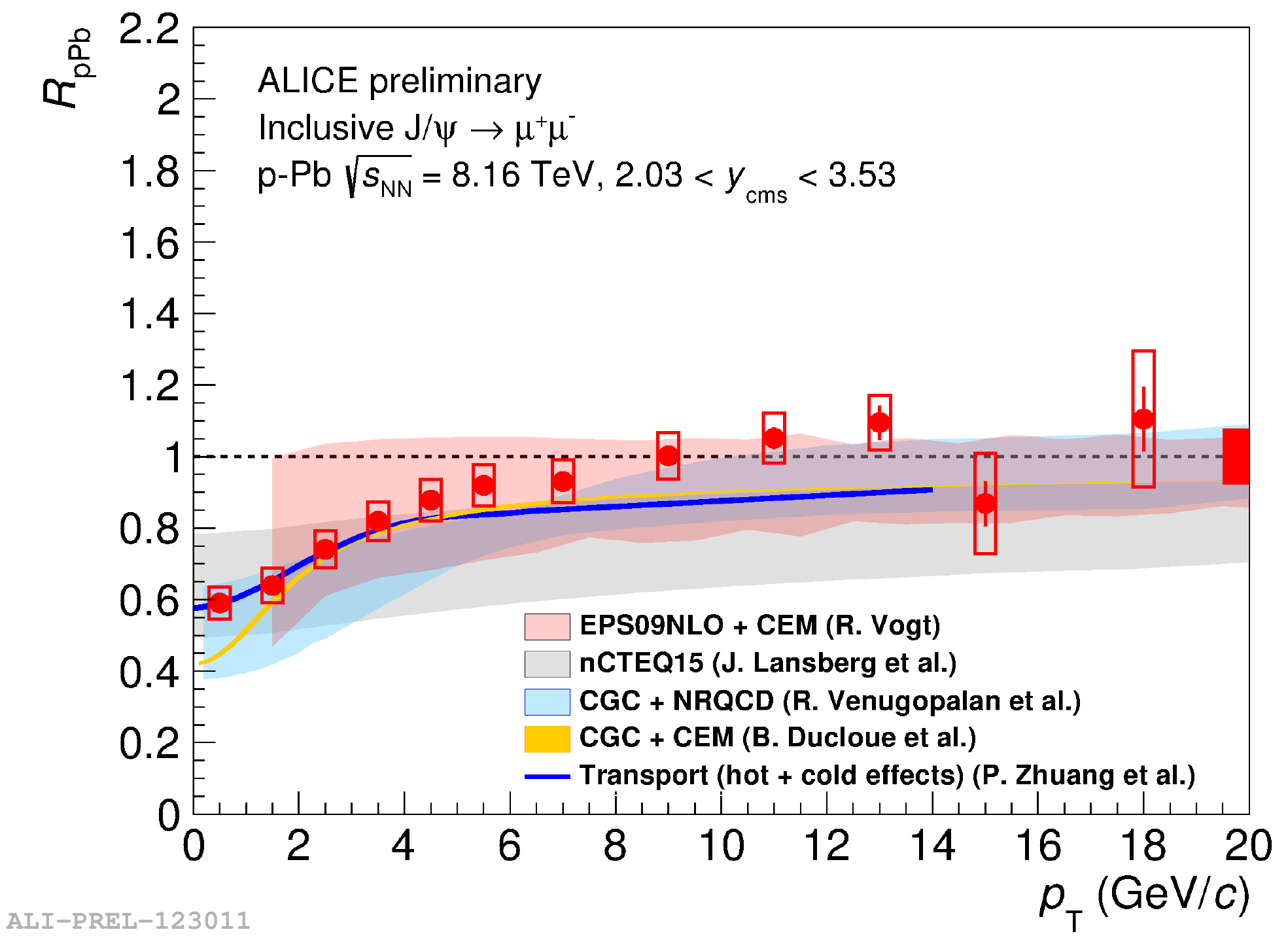} 
\caption{Rapidity (left) and $p_{\rm T}$ (right) dependence of the inclusive J/$\psi$ $R_{\rm pPb}$ for \mbox{p-Pb} collisions at $\sqrt{s_{\rm NN}}=8.16$ TeV, shown by ALICE~\cite{QM17:Tarhini,ALICE-PUBLIC-2017-001}, and compared to theory calculations (see~\cite{ALICE-PUBLIC-2017-001} for the corresponding references).}
\label{fig:4}
\end{center}
\end{figure}

\begin{figure}[hbtp]
\begin{center}
\includegraphics[width=0.48\linewidth]{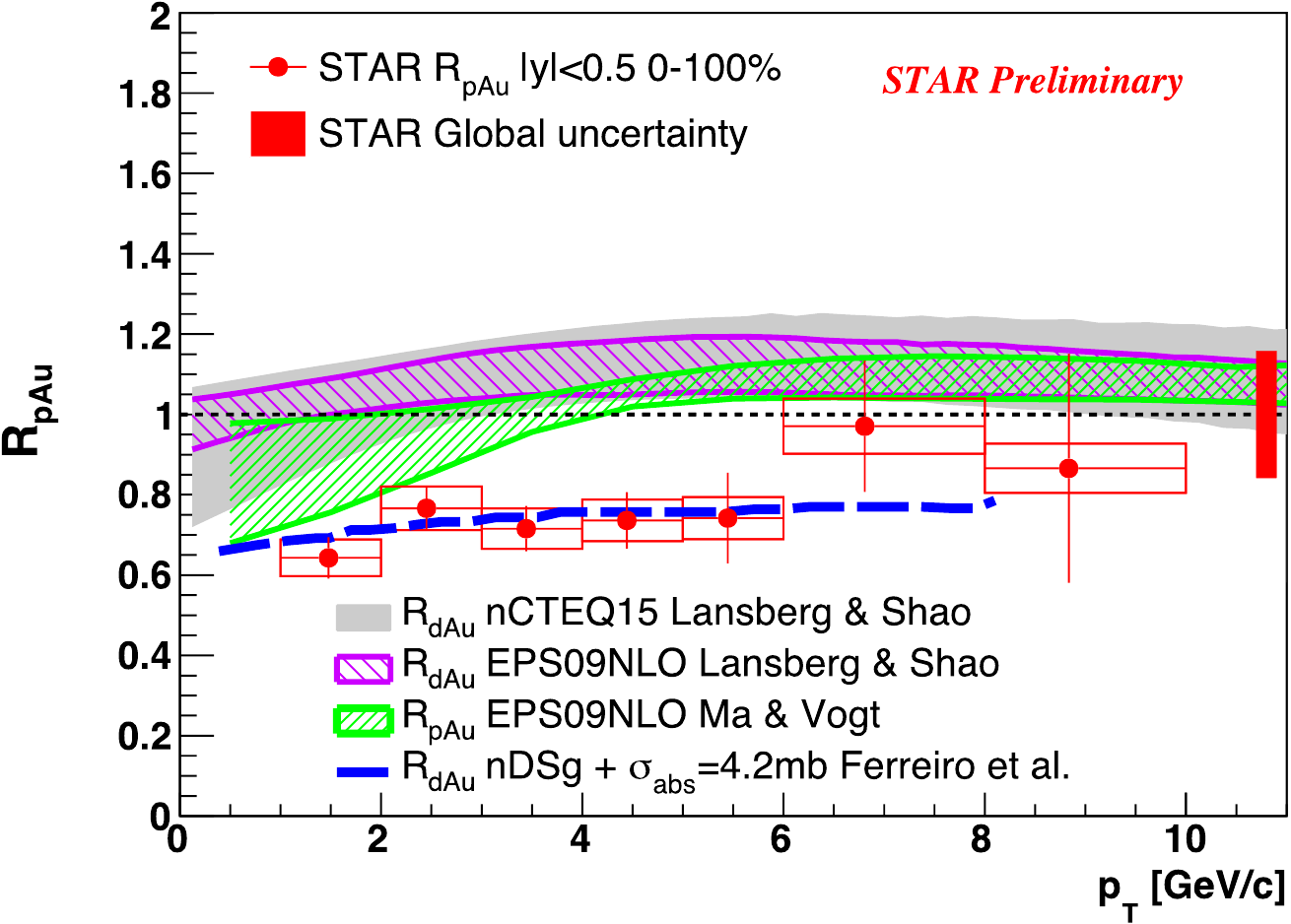} 
\includegraphics[width=0.48\linewidth]{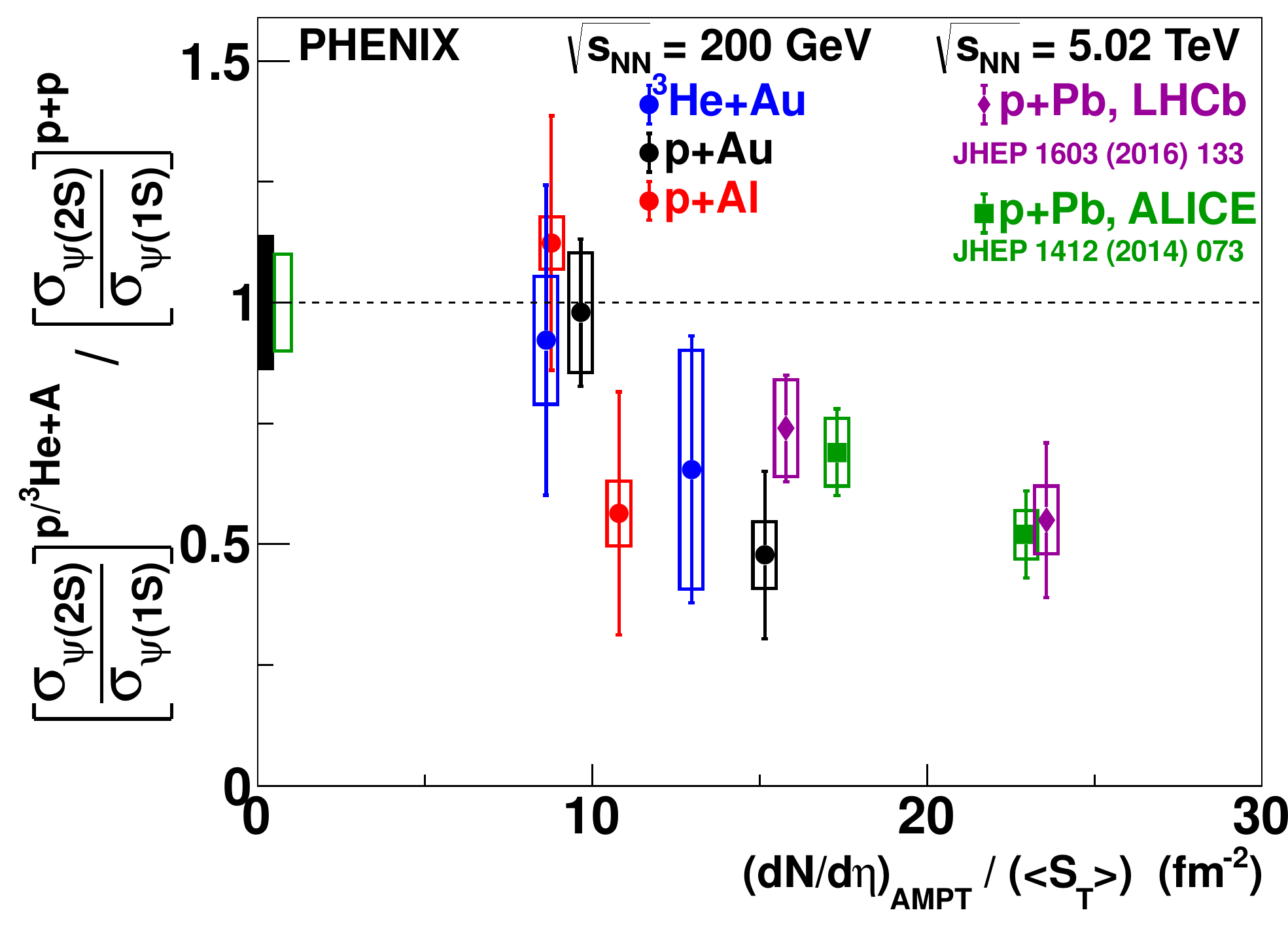} 
\caption{Left: $p_{\rm T}$ dependence of the J/$\psi$ $R_{\rm pAu}$, shown by STAR~\cite{QM17:Todoroki} and compared with the results of theory calculations. Right: double ratio of $\psi(2S)$ and J/$\psi$ cross sections between \mbox{p-A} (or $^3$He--Au) and pp, as a function of the comoving particle density, estimated starting from the collision centrality by using AMPT simulations~\cite{QM17:Lim}.}
\label{fig:5}
\end{center}
\end{figure}

With the very recent availability of high statistics \mbox{p-Pb} data from the LHC run 2 ($\sqrt{s_{\rm NN}}=8.16$ TeV), the possibility of producing high-quality multi-differential results has opened up. For the moment, first results on the J/$\psi$ $R_{\rm pPb}$ were presented by ALICE~\cite{QM17:Tarhini,ALICE-PUBLIC-2017-001} as a function of rapidity and $p_{\rm T}$ (see Fig.~\ref{fig:4}). They are compatible with previous 
results at $\sqrt{s_{\rm NN}}=5.02$ TeV (not shown~\cite{Adam:2015iga}) and are in good agreement with theory models which include various cold nuclear matter effects. In addition, the $p_{\rm T}$ coverage was now extended from 8 up to 20 GeV/$c$.
These results indicate that the effect of J/$\psi$ break-up in nuclear matter
(not included in the models) is weak. Such effect was known to be non-negligible at lower energy, where the $c\overline c$ pair spends more time in nuclear matter. This fact has received another confirmation by preliminary STAR data~\cite{QM17:Todoroki} in \mbox{p-Au} collisions shown at the Conference (Fig.~\ref{fig:5}~(left)), which indicate that only including a J/$\psi$ break-up contribution $\sigma_{\rm abs}=4.2$ mb in the theoretical models a good agreement can be found.

Concerning the $\psi(2S)$, its weaker binding energy may lead to a different behaviour with respect to J/$\psi$. In principle, one would expect at the LHC energies no significant effect from break-up in nuclear matter also for the $\psi(2S)$, because of the short crossing times in the nucleus for the $c \overline c$ pair. However, results at both RHIC and LHC energies~\cite{QM17:Lim,Adare:2016psx,Abelev:2014zpa}, shown in Fig.~\ref{fig:5}~(right), indicate that the double ratio between $\psi(2S)$ and J/$\psi$ in p-Pb and pp is significantly smaller than 1 and decreases for more central collisions. The only explanation for this observation is a suppression effect due to the interaction with the system produced in the collision, which can be a dense hadronic system or even a deconfined medium, if the energy density reached in the collision is large enough. The results from the high statistics \mbox{p-Pb} data from the LHC run 2 will likely help in constraining the models describing this suppression process.

\section{Bottomonium production: A-A collisions}
\label{sec:bottomAA}

In recent years, the study of bottomonia has received considerable attention. The relatively small multiplicity of $b$ quarks, even at the LHC energies, makes the recombination effects rather weak, allowing an easier isolation of phenomena  related to color screening. In addition, the $\Upsilon(1S)$, with a binding energy larger than 1 GeV/$c$, represents a good probe of the very hot QGP formed at the highest available collision energies. Furthermore, the simultaneous study of the 1S, 2S and 3S states, which have increasingly weaker binding energy ($\sim 0.2$ GeV for the 3S) is expected to provide a good sensitivity to the characteristics of the medium. Exploratory studies performed at RHIC~\cite{Adamczyk:2013poh} have given hints for a significant suppressions of the $\Upsilon$ states, while results from the LHC run 1, from ALICE~\cite{Abelev:2014nua} and in particular from CMS~\cite{Khachatryan:2016xxp}, have established a clear hyerarchy in the suppression in \mbox{Pb-Pb} collisions, with $R_{\rm AA}(\Upsilon(1S))\sim 0.45$, the more weakly bound 2S state having $R_{\rm AA}\sim$~0.1, and the 3S state no longer visible.
   
\begin{figure}[hbtp]
\begin{center}
\includegraphics[width=0.48\linewidth]{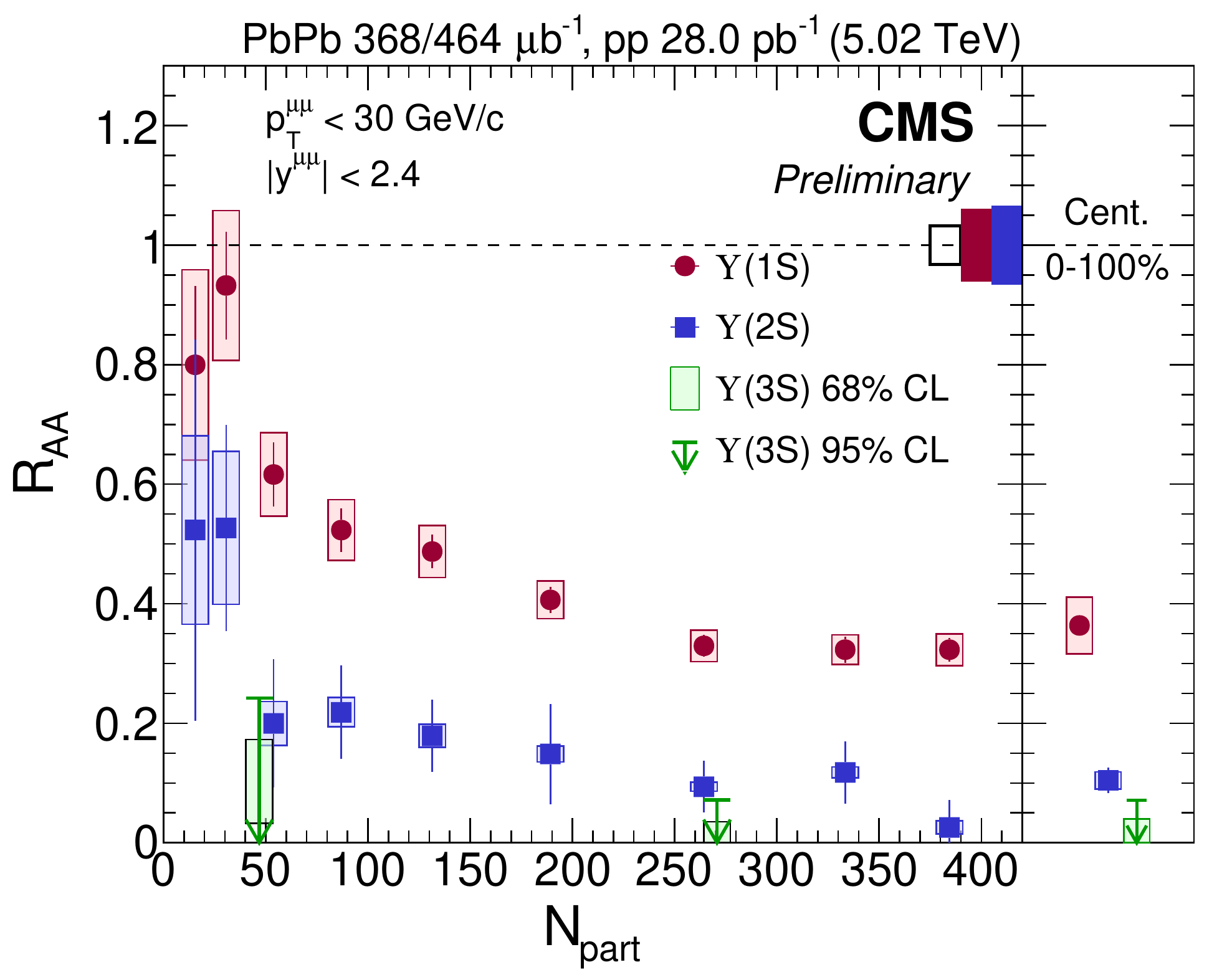} 
\includegraphics[width=0.48\linewidth]{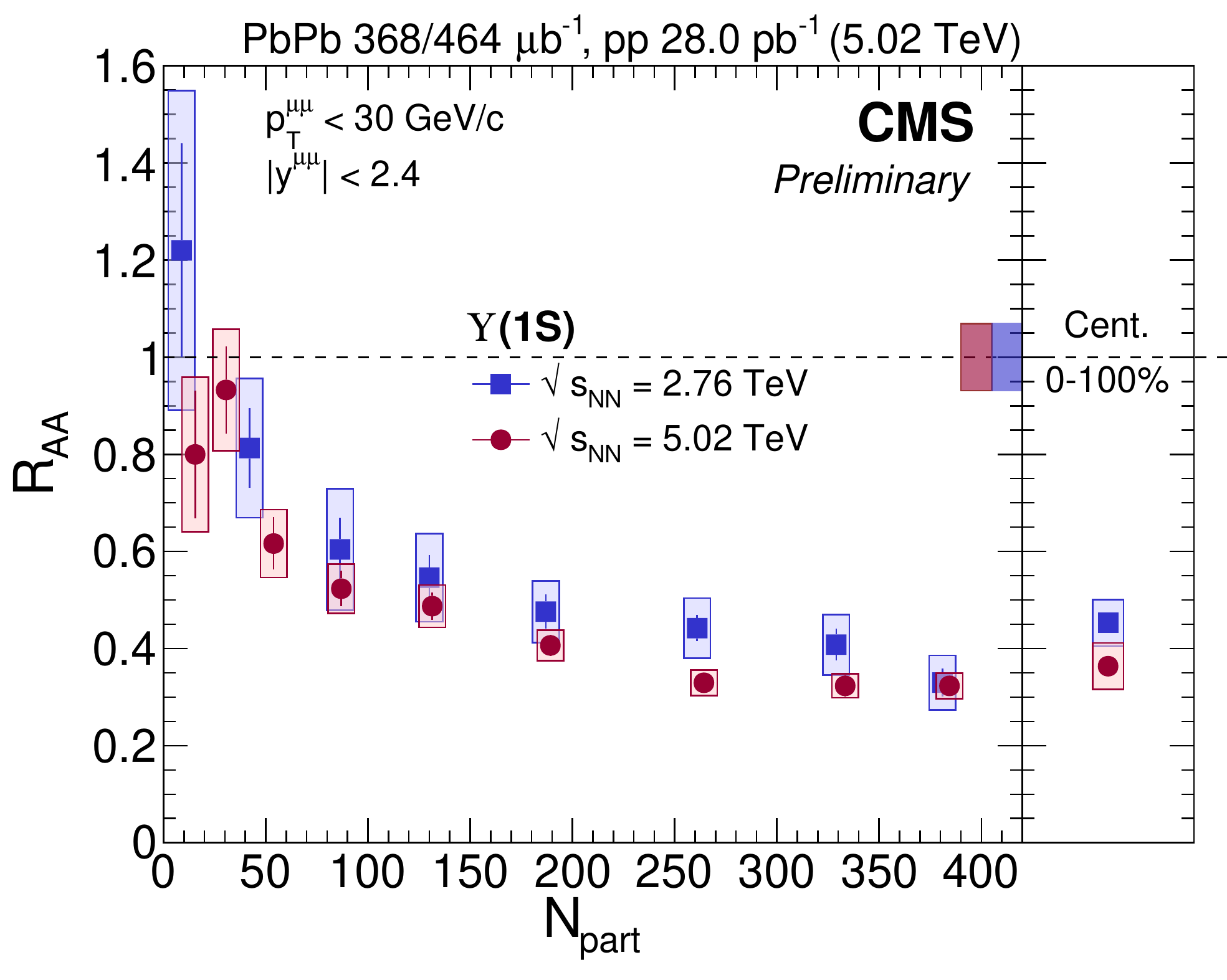} 
\caption{Left: centrality dependence of the $R_{\rm AA}$ for the various $\Upsilon$ resonances in \mbox{Pb-Pb} collisions at $\sqrt{s_{\rm NN}}=5.02$ TeV, shown by CMS~\cite{QM17:Flores,CMS-PAS-HIN-16-023}. Right: comparison of the $\Upsilon(1S)$ $R_{\rm AA}$, as a function of centrality, at $\sqrt{s_{\rm NN}}=5.02$ and 2.76 TeV~\cite{QM17:Flores,CMS-PAS-HIN-16-023}.}
\label{fig:6}
\end{center}
\end{figure}

At this Conference, a certain number of new results from run 2 were presented for the first time. In Fig.~\ref{fig:6}~(left) the centrality dependence of the nuclear modification factor for the various $\Upsilon$ states, measured by CMS~\cite{QM17:Flores,CMS-PAS-HIN-16-023} at $\sqrt{s_{\rm NN}}=5.02$ TeV, is shown, while in Fig.~\ref{fig:6}~(right) a comparison of the $\Upsilon(1S)$ $R_{\rm AA}$ at $\sqrt{s_{\rm NN}}=2.76$ and 5.02 TeV is reported~\cite{QM17:Flores,Khachatryan:2016xxp}. The very strong suppression of the 2S and 3S states confirms the observation carried out by CMS on run 1 data. A slightly stronger suppression at the higher energy is visible for the $\Upsilon(1S)$, but it is not significant with respect to the measurement uncertainties. The observed $R_{\rm AA}\sim 0.3$ for the 1S state in central events points to a strong effect also on this tightly bound resonance. However, it is known that due to the feed-down to the 1S state from higher mass (and more weakly bound) states, it is not straightforward to conclude on the suppression of the directly produced $\Upsilon(1S)$ (see discussion in the next Section).

\begin{figure}[hbtp]
\begin{center}
\includegraphics[width=0.48\linewidth]{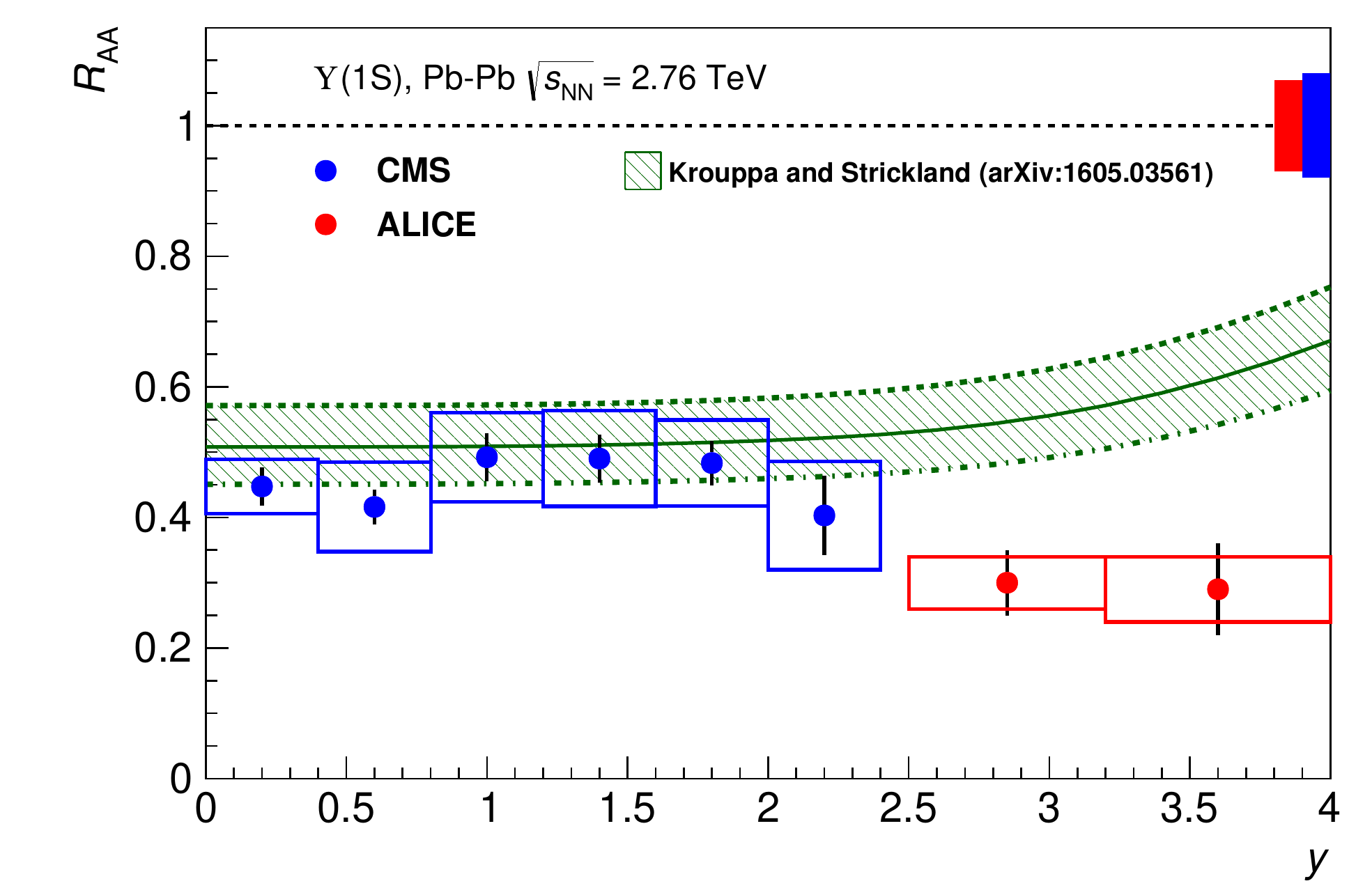} 
\includegraphics[width=0.48\linewidth]{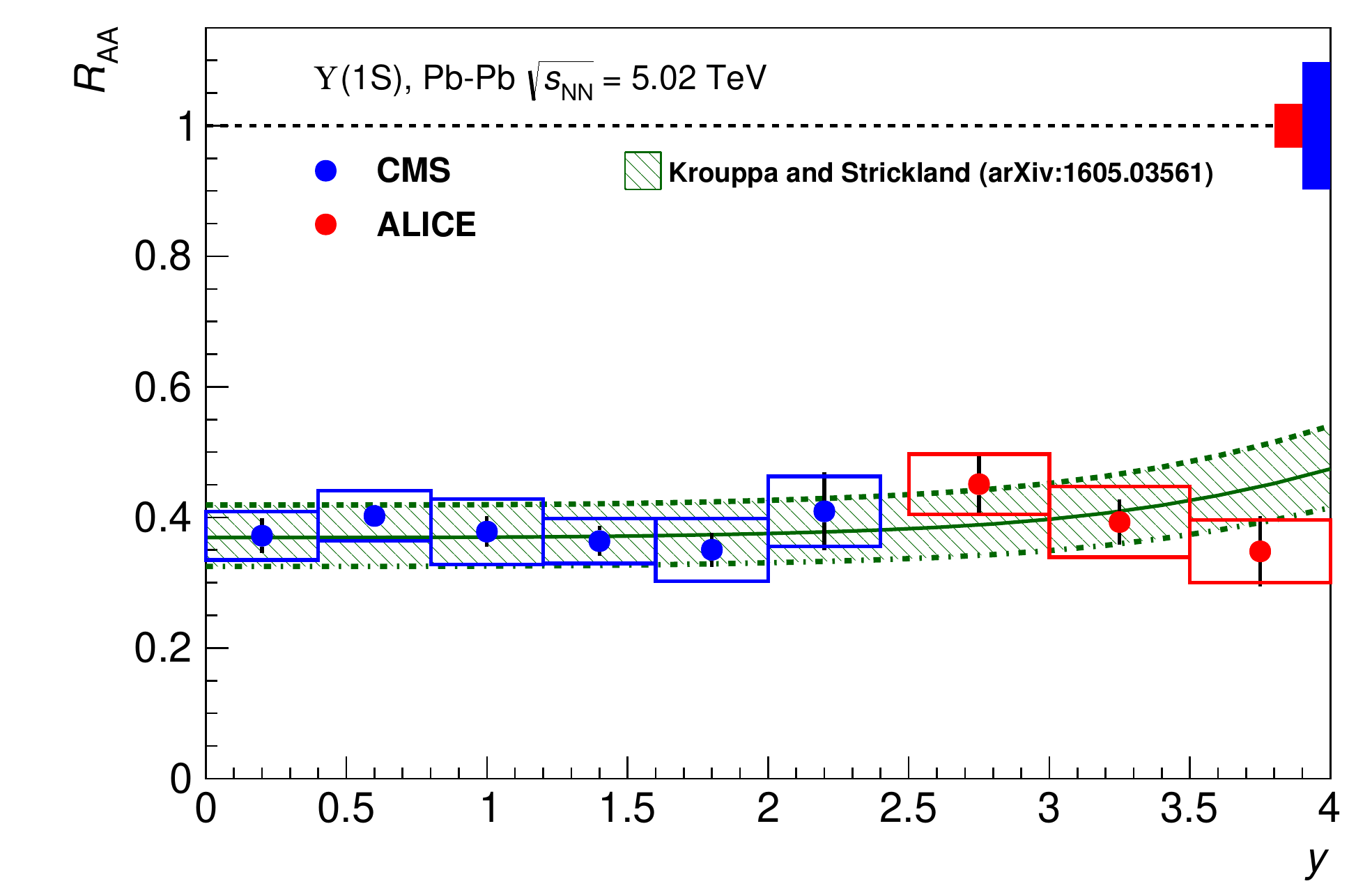} 
\caption{The rapidity dependence of the $\Upsilon(1S)$ $R_{\rm AA}$ at $\sqrt{s_{\rm NN}}=2.76$ (left) and 5.02 TeV (right)~\cite{QM17:Flores,CMS-PAS-HIN-16-023}. Compilation of data from ALICE and CMS, compared with the result of a theoretical model based on an  anisotropic hydrodynamic description of the QGP~\cite{QM17:Das,QM17:Flores,Krouppa:2016jcl}.}
\label{fig:7}
\end{center}
\end{figure}

Considering together bottomonium results from CMS and ALICE, a good coverage in the region $0<y_{\rm cms}<4$ is obtained. In Fig.~\ref{fig:7} the rapidity dependence of the $\Upsilon(1S)$ $R_{\rm AA}$ is shown at the two energies that have been studied at the LHC~\cite{QM17:Flores,QM17:Das}. While at $\sqrt{s_{\rm NN}}=5.02$ TeV an overall flat distribution can be seen, hints for a stronger suppression at large $y_{\rm cms}$ are visible at $\sqrt{s_{\rm NN}}=2.76$ TeV. The latter behaviour is typical in case of the presence of recombination effects, where the higher multiplicity of heavy quarks at central rapidity would induce a stronger regeneration of bottomonia. However, the non-negligible size of systematic uncertainties induces to consider this possibility with caution, also because the recombination effects should in principle be stronger at the higher collision energy. The comparison of the results with a theoretical model based on a hydrodynamical description of the plasma~\cite{Krouppa:2016jcl} shows the opposite effect, with a weaker suppression at large $y$, which clearly leads to a tension with the experimental results at $\sqrt{s_{\rm NN}}=2.76$ TeV.

\begin{figure}[hbtp]
\begin{center}
\includegraphics[width=0.45\linewidth]{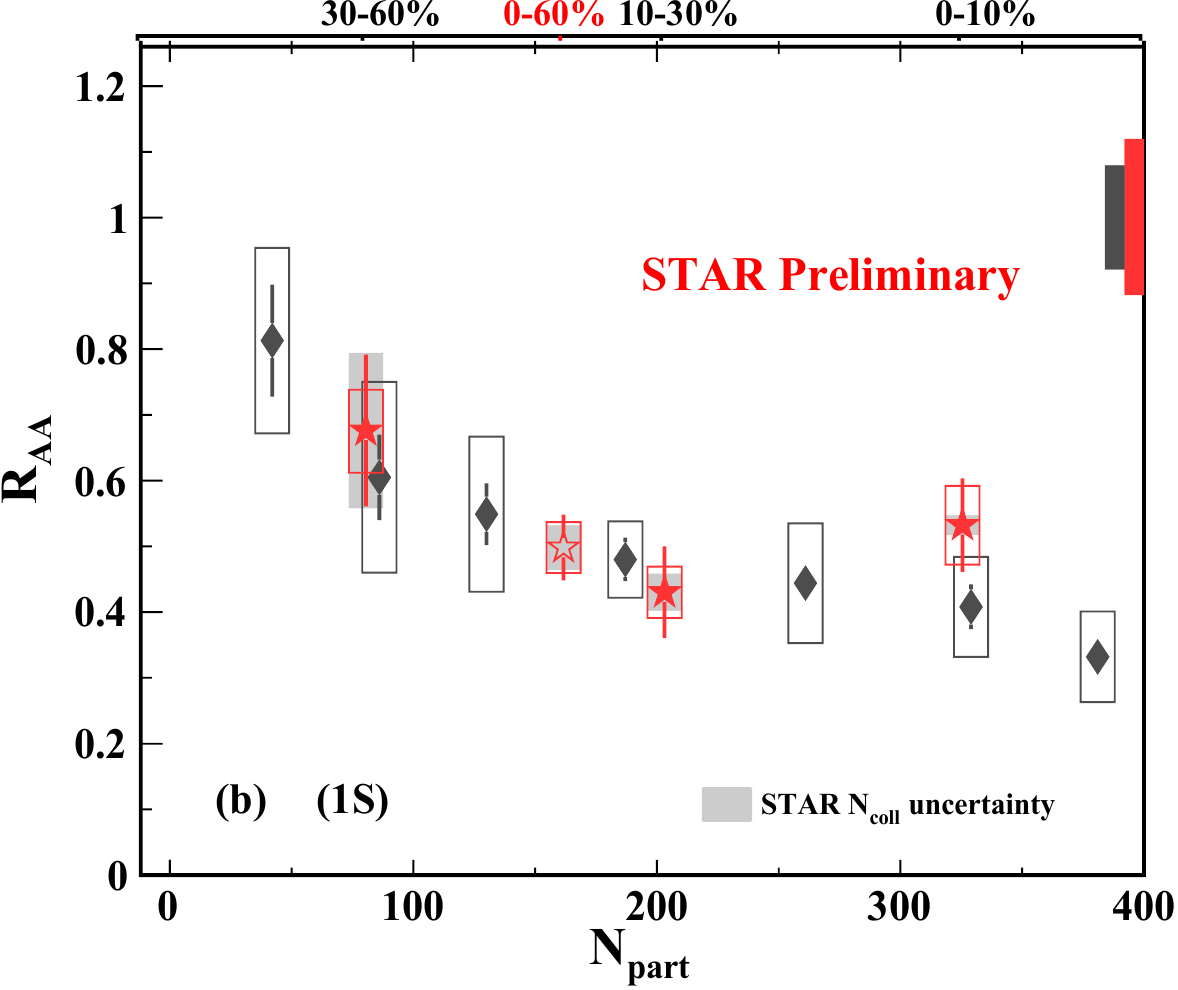} 
\includegraphics[width=0.54\linewidth]{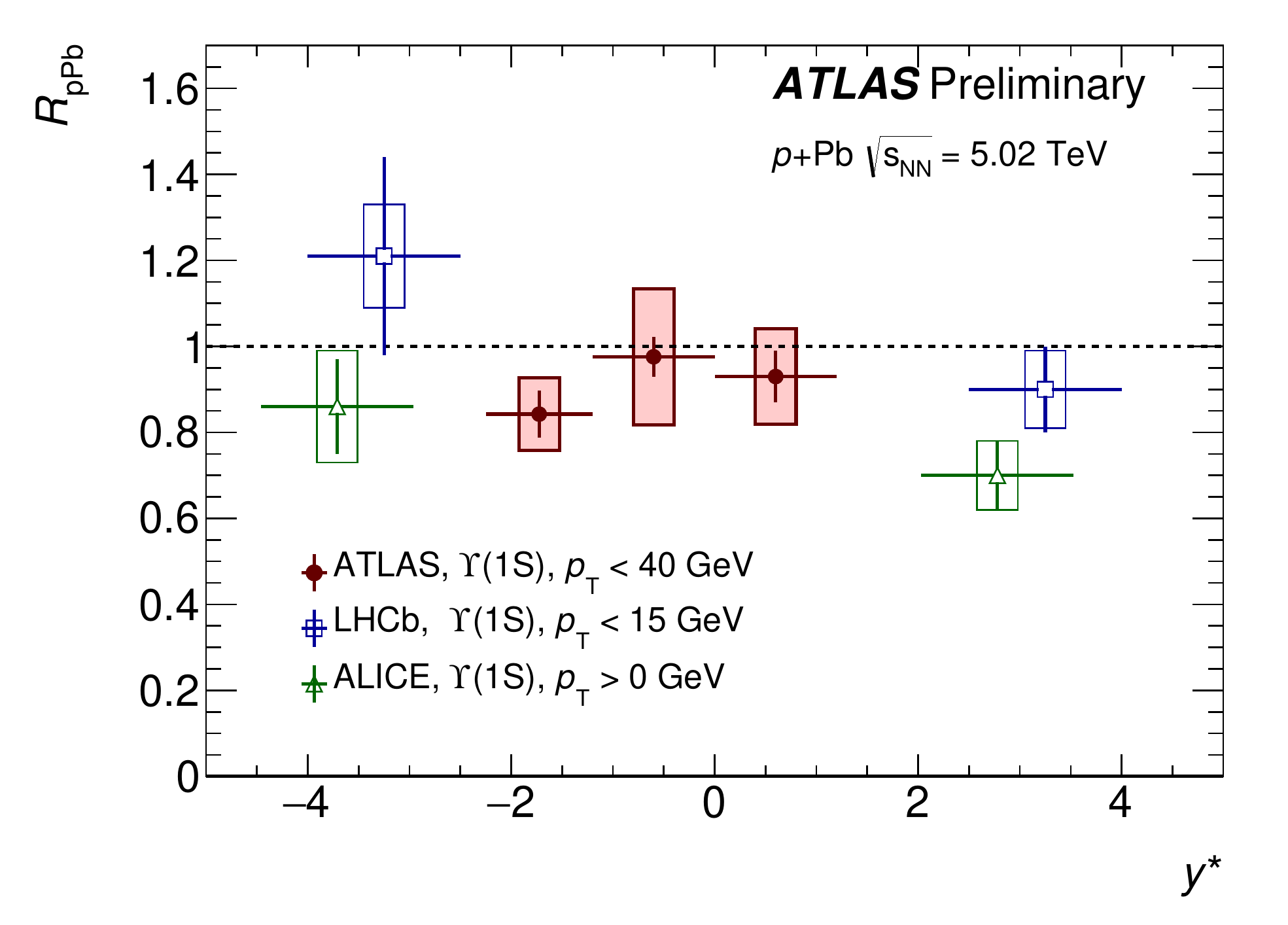} 
\caption{
Left: preliminary results on the centrality dependence of the $\Upsilon(1S)$ $R_{\rm AA}$, measured by STAR in \mbox{Au-Au} collisions at $\sqrt{s_{\rm NN}}=0.2$ TeV~\cite{QM17:Ye} and compared with the corresponding CMS results  (stars)~\cite{Khachatryan:2016xxp}.
Right: rapidity dependence of $R_{\rm pPb}$ for $\Upsilon(1S)$ in \mbox{p-Pb} collisions at $\sqrt{s_{\rm NN}}=5.02$ TeV, with results from ATLAS, LHCb and ALICE~\cite{ATLAS-CONF-2015-050,Aaij:2014mza,Abelev:2014oea}.}
\label{fig:8}
\end{center}
\end{figure}

In addition to the rich sample of results from the LHC experiments, new bottomonium results from \mbox{Au-Au} collisions at $\sqrt{s_{\rm NN}}=0.2$ TeV were presented by STAR at the Conference, separately for the $\Upsilon(1S)$ and for the sum of the 2S and 3S states~\cite{QM17:Ye}.
For the $\Upsilon(2S+3S)$, the $R_{\rm AA}$, compared to CMS results for the two resonances~\cite{Khachatryan:2016xxp}, shows an indication for a weaker suppression at RHIC energies up to $N_{\rm part}\sim 200$, while for central collisions the suppression effects become quite similar. Qualitatively, this observation could be connected with the increase of the energy density with centrality, followed by a saturation of the suppression effect from central collisions at RHIC up to LHC energy. For the $\Upsilon(1S)$ (Fig.~\ref{fig:8}~(left)), the centrality dependence of $R_{\rm AA}$ is very similar for STAR and CMS results, an observation which could be explained if the suppression were mainly due to the feed-down effect from $\Upsilon(2S)$ and $\Upsilon(3S)$, with little or no effect on the direct $\Upsilon(1S)$ at both RHIC and LHC. However, to confirm this scenario, an accurate knowledge of cold nuclear matter effects and of feed-down contributions from higher mass resonances, together with quantitative model comparisons, are necessary.

\section{Bottomonium production: p-A collisions}
\label{sec:bottompA}

Similarly to the charmonium case, also for bottomonia cold nuclear matter effects can in principle be sizeable, and need to be estimated via proton-nucleus collisions. For the moment, only the LHC run 1 results are available and their quality, as can be seen in Fig.~\ref{fig:8}~(right) is still lower than for charmonia. In particular, ATLAS mid-rapidity results in \mbox{p-Pb} collisions at $\sqrt{s_{\rm NN}}=5.02$ TeV~\cite{ATLAS-CONF-2015-050} favour $R_{\rm pPb}$ values close to 1, while at forward (p-going) and backward (Pb-going) rapidity, LHCb~\cite{Aaij:2014mza} and ALICE~\cite{Abelev:2014oea} results, although compatible, span a rather large range of possible values. The trend from small (or no) cold nuclear matter effects at negative $y$ towards a suppression at positive $y$, clearly observed for J/$\psi$, seems visible also for the $\Upsilon(1S)$. Comparisons with theory models (not shown) are still not conclusive, due to the large uncertainties of the experimental results. From a qualitative point of view, one may attempt to make use of \mbox{p-Pb} data to evaluate the presence of a suppression signal for directly produced $\Upsilon(1S)$ in \mbox{Pb-Pb}. Restricting these considerations to the region around mid-rapidity, the ATLAS \mbox{p-Pb} data allow, at 1$\sigma$ level, $R_{\rm pPb}\sim 0.8$, which, assuming a dominance of shadowing effects, would imply for \mbox{Pb-Pb} collisions, $R_{\rm PbPb}^{\rm CNM}\sim 0.8^2=0.64$. On the other hand, CMS measures, in \mbox{Pb-Pb} collisions at the same energy, for inclusive $\Upsilon(1S)$ production, $R_{\rm PbPb}^{\rm incl} = 0.36$~\cite{CMS-PAS-HIN-16-023}. Recent estimates of feed-down effects, based mostly on LHCb data (see e.g.,~\cite{Aaij:2012se}), indicate a $\sim30$\% effect, which would lead to 
$R_{\rm PbPb}^{\rm direct} \sim 0.36/0.7 = 0.51$, a value smaller than CNM extrapolations. Even if this evaluation is admittedly qualitative, the conclusion that can be drawn is that we are likely observing a suppression of direct $\Upsilon(1S)$, which, according to lattice calculations, would imply a QGP temperature at least $~\sim 2 T_{\rm c}$ at the LHC energies~\cite{Burnier:2015tda}.

Finally, also for $\Upsilon(2S)$ and $\Upsilon(3S)$ no precise evaluations of $R_{\rm pPb}$ are available for the moment. A relative measurement (double ratio between the excited and the 1S state between \mbox{p-Pb} and pp) was published by CMS~\cite{Chatrchyan:2013nza}. It clearly indicates that in \mbox{p-Pb} collisions the 2S and 3S states are more suppressed than the $\Upsilon(1S)$. As for the $\psi(2S)$ vs J/$\psi$ case, one does not expect a different $R_{\rm pPb}$ between the various states due to either shadowing or interactions with nuclear matter, and the only mechanism that may explain this observation is the break-up of the 2S and 3S states via interactions with the medium (either partonic or hadronic) produced in the \mbox{p-Pb} collision.


\section{Conclusions}
\label{sec:conclusions}

After many years, the study of quarkonium production continues to be a source of new and relevant results for our understanding of the QGP created in heavy-ion collisions. Several highlights were presented at Quark Matter 2017: for charmonium production the first unambiguous observation of a strong elliptic flow for the J/$\psi$ at low $p_{\rm T}$ has been established, and precise new data on J/$\psi$ suppression and regeneration have become available, thanks to the first analyses of the LHC run 2 data in \mbox{Pb-Pb} collisions. Also the $\psi(2S)$ results are becoming more accurate, and even if their $\sqrt{s_{\rm NN}}$-dependence is still not completetly understood, they offer the possibility of a deeper insight on charmonium properties in the medium.
Concerning bottomonia, detailed information on $\Upsilon(1S)$ and $\Upsilon(2S)$ $R_{\rm AA}$ is now available at both $\sqrt{s_{\rm NN}}=2.76$ and 5.02 TeV, and  an unambiguous evidence for a hierarchy of suppression was reached. The description of the $y$-dependence of $R_{\rm AA}$ is still not fully satisfactory, and might need some further refinement in the models. Finally, a first set of precise results from RHIC has become available, and will surely represent an important ingredient towards a unified description of the bottomonium phenomenology from low to high energy.    




\bibliographystyle{elsarticle-num}
\bibliography{Scomparin_QM17_short}







\end{document}